\documentclass[useAMS]{mn2e}
\usepackage{psfig}
\usepackage{times}

\def\ltsima{$\; \buildrel < \over \sim \;$}
\def\lsim{\lower.5ex\hbox{\ltsima}}
\def\gtsima{$\; \buildrel > \over \sim \;$}
\def\gsim{\lower.5ex\hbox{\gtsima}}

\def\bb{black--body}

\title[Re--born fireballs] 
{Re--born fireballs in Gamma-Ray Bursts}

\author[G. Ghisellini et al.]
{G. Ghisellini$^1$, A. Celotti$^2$, G. Ghirlanda$^1$, C. Firmani$^{1,3}$, L. Nava$^{1,4}$\\
$^{1}$INAF -- Osservatorio Astronomico di Brera, via E. Bianchi 46, I--23807 Merate, Italy\\
$^{2}$SISSA, via Beirut 2--4, I--34014 Trieste, Italy  \\
$^{3}$Instituto de Astronom\'ia, U.N.A.M., A.P. 70-264, 04510, M\'exico, D.F., M\'exico\\
$^{4}$Universit\'a degli Studi dell'Insubria, Dipartimento di 
Fisica e Matematica, via Valleggio 11, I--22100 Como, Italy
}

\begin{document}
\maketitle
\label{firstpage}

\begin{abstract}
We consider the interaction between a relativistic fireball and
material assumed to be still located just outside the progenitor star.
Only a small fraction of the expected mass is sufficient to
efficiently decelerate the fireball, leading to dissipation of most of
its kinetic energy.  Since the scattering optical depths are still
large at distances comparable to the progenitor radius, the dissipated
energy is trapped in the system, accelerating it to relativistic
velocities.  The process resembles the birth of another fireball at
radii $R\sim 10^{11}$ cm, not far from the transparency radius, and
with a starting bulk Lorentz factors $\Gamma_{\rm c}\sim 10$.  As seen
in the observer frame, this ``re--generated" fireball appears
collimated within an angle $\theta_{\rm j} = 1/\Gamma_{\rm c}$.  If
the central engine works intermittently, the funnel can, at least
partially, refill and the process can repeat itself.  We discuss how
this idea can help solving some open issues of the more conventional
internal shock scenario for interpreting the Gamma--Ray Burst
properties.
\end{abstract}


\section{Introduction}

The internal/external shock scenario (see e.g. Piran 2004; Meszaros
2006) is currently the leading model to explain the complex
phenomenology of Gamma--Ray Bursts (GRBs) prompt and afterglow
emission.  Despite it can account for many observed characteristics,
there are a few open issues and difficulties that this model cannot
solve, or can accommodate only with some important modifications.

Here we recall some problems of the standard scenario and mention some
ideas already put forward to account for them.

\begin{itemize}
\item
{\it Efficiency I: high $\Gamma$--contrast ---} In internal
shocks only the relative kinetic energy of the two
colliding  shells can be dissipated. 
Thus ``dynamical" efficiencies of only a few per cent can
be achieved for colliding shells whose Lorentz factors $\Gamma$ differ
by a factor of order unity.  Such efficiency has to include energy
dissipated into randomizing protons, amplifying (or even generating)
magnetic fields and accelerating emitting leptons.  As the emitted
radiation is produced only by the latter component, it corresponds to
just a fraction of the dynamical efficiency.  This problem, pointed
out, among others, by Kumar (1999) can be solved by postulating
contrasts in $\Gamma$ much exceeding 100 (Belobodorov 2000; Kobayashi
\& Sari 2001).  In these cases the typical Lorentz factor of GRBs
should thus largely exceed the ``canonical" value $\sim$ 100. 
In this case it is difficult to understand how the value
of the peak energy of the prompt spectrum does not wildly change.

\item
{\it Efficiency II: Afterglow/Prompt power ratio ---} A related
inconsistency concerns the observed ratios of the bolometric fluence
originated in the afterglow to that in the prompt phase.  Since
external shocks are dynamically more efficient than internal ones,
such ratio is expected to exceed one, contrary to current estimates.
The problem has been exacerbated by the recent observations by the
Swift satellite, showing that the X--ray afterglow light curve seen
after a few hours -- thought to be smoothly connected with the end of
the prompt -- comprises a steep early phase.  As a consequence, the
total afterglow energy is less than what postulated before.
Willingale et al. (2007), parameterizing the behavior of the Swift GRB
X--ray light curves, derived an average X--ray afterglow--to--prompt
fluence ratio around 10 per cent 
(see also Zhang et al. 2007).
Furthermore, the very same origin
of the early X--ray radiation as produced by external shocks is
questioned, since its behavior is different from the optical one
(e.g. Panaitescu et al. 2006).  If the X--ray emission does not
originate in the afterglow phase, this further reduces the above
ratio.

\item
{\it The spectral energy correlations ---} Correlations have been
found between i) the energy where most of the prompt power is emitted
$(E_{\rm peak})$ and the isotropic prompt bolometric energetics
$E_{\rm \gamma, iso}$ (Amati et al. 2002; Amati 2006), and 
ii) $E_{\rm peak}$ and the collimation corrected energetics $E_{\gamma}$.  
The slope of the former correlation is $E_{\rm peak}\propto 
E_{\rm \gamma, iso}^{1/2}$ while the slope of the latter depends on the radial
profile of the circumburst density.  For a homogeneous density medium
$E_{\rm peak}\propto E_{\rm \gamma}^{0.7}$ (Ghirlanda et al. 2004),
while for a wind--like profile in density ($\propto r^{-2}$) the
correlation is linear: $E_{\rm peak}\propto E_{\rm \gamma}$ (Nava et
al. 2006; Ghirlanda et al. 2007a).  If linear, the relation is Lorentz
invariant and indicates that different GBRs roughly emit the same
number of photons at the peak (i.e. $E_\gamma/E_{\rm peak} \sim$
constant).

Note that the derivation of $E_\gamma$ requires not only information
on the jet break time $t_{\rm j}$, but also a model relating $t_{\rm
j}$ with the collimation angle $\theta_j$, which in turn depends on
the circumburst density value, profile and the radiative efficiency
$\eta$ (i.e. $E_{\gamma} =\eta E_{\rm kin}$).  The phenomenological
connection among the three observables $E_{\rm \gamma, iso}$, $E_{\rm
peak}$ and $t_{\rm j}$, as found by Liang \& Zhang (2005), is instead
model--independent.  It is of the form $E_{\rm iso} \propto E_{\rm
peak}^a t_{\rm j}^{-b}$, which for $b\sim -1$ is consistent with the
Ghirlanda relation (both in the homogeneous and wind case; see Nava et
al. 2006).  A further tight phenomenological relation appears to link
three prompt emission quantities: the isotropic peak luminosity
$L_{\rm iso}$, $E_{\rm peak}$ and the time interval $T_{0.45}$ during
which the emission is above a certain level (Firmani et al. 2006). 
All these correlations were not predicted by the
internal/external shock scenario, and can only be reconciled with it
as long as specific dependences of the bulk Lorentz factor upon 
$E_{\rm iso}$ are satisfied (see Table 1 in Zhang \& Meszaros 2002).

\end{itemize}

The above issues motivate the search for alternatives or 
for substantial modifications of the standard model.  
The efficiency problem and the existence of
the spectral--energy relations prompted Thompson (2006, T06 hereafter)
and Thompson Meszaros \& Rees (2007, T07 hereafter) to suggest that,
besides internal shocks, dissipation might also occur because of the
interaction between the fireball and the walls of the funnel in the
star through which it propagates (see \S 2).  This hypothesis also
introduces a typical scale to the problem, namely the radius of the
progenitor star ($R_*\sim 10^{10}$--$10^{11}$ cm): shear instabilities
within $R_*$ can reconvert a significant fraction of bulk kinetic
energy into heat.  Since this dissipation occurs up to $R_*$ (i.e. not
far from the transparency radius), the increased internal energy can
only partially reconvert into bulk motion via adiabatic expansion,
increasing the efficiency. Similarly, studies of magnetized fireballs
(e.g. Drenkhahn \& Spruit, 2002; Giannios \& Spruit 2007) have shown
that dissipation of magnetic energy through reconnection can also
contribute to increase the radiation content of the fireball at
relatively large radii.

Along the above mentioned lines, in this Letter we propose a further
possible way in which a large fraction of the fireball bulk energy can
be dissipated at distances $R\sim R_*$. This assumes that at these
distances the fireball collides with some mass which is (nearly) at
rest: a small fraction of the mass swept up in the funnel left along
the fireball propagation axis is sufficient to lead to efficient
dissipation. 
Hereafter such mass will be referred to as ``IDM" (Intervening Debris
of the cocoon Material).

The treatment of the collision is simplified, in order to allow an
analytical and simple description. We assume the IDM  to be at
rest and homogeneous in density and the fireball to have a Lorentz
factor $\Gamma \gg 1$. The interaction is described in ``steps", while
in reality will be continuous in time.  A complete treatment of the
dynamics and emission properties of our model requires numerical
simulations (of the kind presented in Morsony et al. 2007, introducing
some erratic behavior of the injected jet energy). Interestingly, the
model predicts that jet properties depend on the polar angle (like in
a structured jet, see Rossi et al. 2002) and it naturally implies a 
connection among the observed spectral--energy correlations.

\section{Shear--driven instabilities and dissipation of bulk kinetic energy}

As mentioned, T06 and T07 proposed a model in which the efficiency of
the dissipation of kinetic energy into radiation is enhanced with
respect to the internal shock scenario and the spectral energy
correlations, in particular the Amati one, can be accounted for.  At
the same time, in their scenario synchrotron emission could play a
minor role, the radiation field being dominated by thermalized high
energy photons or by the inverse Compton process.

For what follows it is useful to summarize their main arguments here.
Consider a fireball that at some distance $R_0$\ltsima$R_*$ from the
central engine is moving relativistically with a bulk Lorentz factor
$\Gamma_0$.  The fireball is initially propagating inside the funnel
of the progenitor star.  T06 and T07 assume that a large fraction of
the energy dissipated at $R_0$ is thermalized into \bb\ radiation of
luminosity
\begin{equation}
L_{\rm BB, iso}\, =\, 4\pi R_0^2 \Gamma_0^2 \sigma {T'_0}^4\, = \,
4\pi {R_0^2\over \Gamma_0^2} \sigma T_0^4, 
\label{bb1}
\end{equation}
where $T'_0$ and $T_0=\Gamma_0 T'_0$ are the temperatures at $R_0$ in
the comoving and observing frame, respectively.  
The collimation corrected luminosity is
$L_{\rm BB} =(1-\cos\theta_{\rm j})L_{\rm BB, iso}$ which, for small 
semi--aperture angles $\theta_{\rm j}$ of the
jetted fireball (assumed conical), gives
\begin{equation}
\theta_{\rm j}^2 \, \sim \, { 2 L_{\rm BB}\over L_{\rm BB, iso}}.
\end{equation}
A key assumption of the model is that 
$\Gamma_0 \sim 1/\theta_{\rm j}$.  
The argument behind it is that if $\Gamma_0>> 1/\theta_{\rm j}$,
shear--driven instabilities have not time to grow (in the comoving
frame), while in the opposite case the flow mixes easily with the
heavier material and decelerates to $\Gamma\sim 1$.  Then, assuming
$\Gamma_0 =1/\theta_{\rm j}$ and substituting it in Eq. \ref{bb1}
\begin{equation}
L_{\rm BB, iso}\, \sim \, 8 \pi R_0^2 { L_{\rm BB}\over 
L_{\rm BB, iso} } \sigma {T_0}^4.
\end{equation}
\label{bb2}
Setting $E_{\rm BB, iso}=L_{\rm BB, iso} t_{\rm burst}$ and 
$E_{\rm BB}=L_{\rm BB} t_{\rm burst}$, where $t_{\rm burst}$ is the duration
of the prompt emission, gives
\begin{equation}
E_{\rm peak}\, \propto\, T_0\, \propto\, E_{\rm BB, iso}^{1/2} E_{\rm
BB}^{-1/4} t_{\rm burst}^{-1/4}.
\label{bb3}
\end{equation}
This corresponds to the Amati relation if $E_{\rm BB}$ is similar in
different bursts and the dispersion of GRB durations is also
limited. It should be noted that a relation similar to the Amati one
can be also recovered adopting $E_{\rm BB}\propto E_{\rm peak}^a$, as
suggested by the Ghirlanda relation. For instance, for $a=1$ (wind case)
\begin{equation}
E_{\rm peak}\, \propto\,  
E_{\rm BB, iso}^{2/5}  t_{\rm burst}^{-1/5}.
\end{equation}
For the derivation of Eq. 4 (and 5) a key assumption is the dependence
on temperature of the black--body law, which leads to both a slope
{\it and} a normalization similar to those characterizing the Amati
relation.

We can ask what happens if, instead of a black--body, one
assumes that the spectrum is a cutoff power--law.
This question is particularly relevant since 
the burst spectrum is rarely described by a pure black--body
(even if some bursts is, see Ghirlanda et al. 2003),
and also the black--body plus power--law model (Ryde 2005)
faces severe problems,
even considering time resolved spectra (Ghirlanda et al. 2007b).

How is the above derivation modified if, instead of a black--body emission,
the spectrum is best described by a cutoff power--law?  
Consider then a spectrum described in the comoving frame by
$L_{\rm \gamma, iso}^\prime(E^\prime) \propto
E^{'-\beta} \exp(-E^\prime/E^\prime_0)$ 
and approximate the observed isotropic 
bolometric luminosity as 
\begin{equation}
L_{\rm iso} \, \propto \, \Gamma_0^2 \left( {E_{\rm peak} \over \Gamma_0 } 
\right)^{1-\beta}
\propto \left( {L_\gamma}  \over L_{\rm \gamma, iso} \right)^{-(1+\beta)/2}
  E_{\rm peak}^{1-\beta},  
\end{equation}
This leads to
\begin{equation}
E_{\rm peak}\, \propto \, E_{\rm iso}^{1/2}
E_\gamma^{(1+\beta)/(2-2\beta)} t_{\rm burst}^{-1/(1-\beta)}.
\label{cpl}
\end{equation}
Therefore the dependence of the Amati relation can be recovered even
for cutoff power--law spectra, but the normalization in this case is
not determined.  Note also that for $\beta=-3$ (Wien spectrum),
Eq. \ref{cpl} and Eq. \ref{bb3} have the same dependences.

\section{Fireball--IDM collision}

To excavate a funnel inside a progenitor star of mass $M_*=10 M_{*,1}$
solar masses, the ``proto--jet" has to push out 
the mass that did not fall into the newly born black hole.
This means a fraction of
$(1-\cos\theta_{\rm f}) M_* = 0.1 M_{*,1} \theta^2_{\rm f, -1}$, where
$\theta_{\rm f} =0.1 \theta_{f,-1}$ is the funnel opening angle.  
This mass expands sideways as the proto--jet breaks out at the
surface of the progenitor, forming a cocoon (see also Ramirez-Ruiz,
Celotti \& Rees 2002).  We must expect, however, that after the
break--out the region in front of the funnel will not be perfectly
cleared of mass.  To be irrelevant, the mass $M_{\rm c}$ left as IDM
should be $\ll E_0/(\Gamma_0 c^2) \sim 5\times 10^{-7} M_\odot$: less
than one in a million particles should remain there. 

If the jetted fireball is not continuous, this mass may be still there
at the moment of the arrival of the new fireball pulse.  And even if
$M_{\rm c}$ is $(10^{-3}$--$10^{-4})$ the excavated mass, the IDM can
have important dynamical effects on it.  As the bulk velocity and
energy content of the IDM can be neglected in comparison to those of
the coming fireball, it will be approximated as initially at rest and cold.

An interesting aspect of this scenario concerns multi--peaked bursts,
especially when pulses in the prompt emission are separated by
quiescent periods.  If the central engine works at a reduced rate
during quiescence, material from the walls of the funnel, previously
in pressure equilibrium with the jet, will tend to refill again the
funnel (see also Wang \& Meszaros 2007) .  
This requires a time $T_{\rm f} \sim \theta_{\rm f} R/
(\beta_{\rm s} c) \sim 3\times 10^{-4} (R/R_{0, 7})\theta_{\rm
f,-1}/\beta_{\rm s, -1}$ s, where $\beta_{\rm s} c$ is the sound
speed, and $R_0=10^7 R_{0,7}$ cm is the radius at the base of the
funnel.  Then the amount of mass which should be pushed out again
depends upon the quiescent time, but that after $\sim$1 second, the
funnel is completely closed, and the process repeats itself.

\subsection{Results: dynamics and dissipation}

In the following we derive the main characteristics of the system
formed by a fireball impacting against the IDM.  Consider a fireball with
energy $E_0$, mass $M_0$ and bulk Lorentz factor $\Gamma_0=E_0/(M_0
c^2)$, impacting against a mass $M_{\rm c}$, initially at rest.

\begin{figure}
\vskip -0.7 true cm
\centerline{\psfig{figure=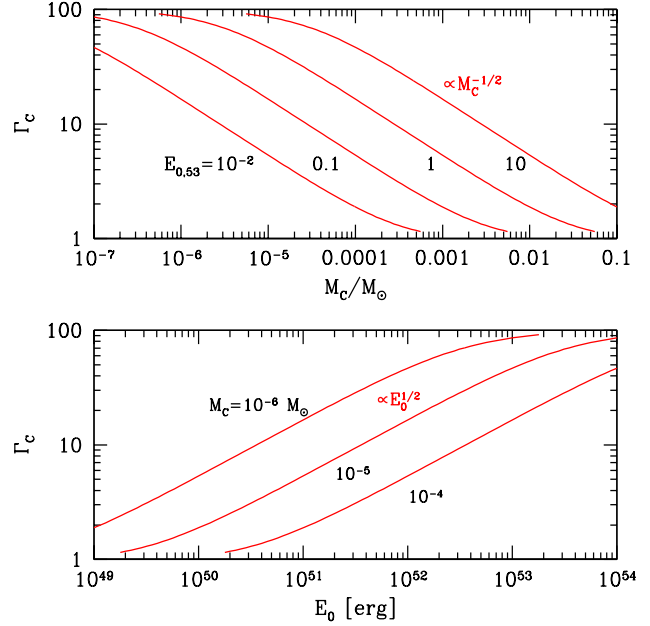,width=9.5cm,height=10cm}}
\vskip -1 true cm
\caption{The Lorentz factor $\Gamma_{\rm c}$ of the IDM+fireball
system as a function of $M_{\rm c}$ for different values of $E_0$ (top
panel) and as a function of $E_0$ for selected values of $M_{\rm c}$
(bottom panel).}
\label{tappo1}
\end{figure}

The energy and momentum conservation laws read
\begin{eqnarray}
M_0 \Gamma_0 + M_{\rm c} \, &=&\,
 \Gamma_{\rm c} (M_0 + M_{\rm c} + \epsilon^\prime/c^2) \nonumber \\ 
M_0 \Gamma_0\beta_0 \, &=&\,
 \Gamma_{\rm c}\beta_{\rm c} (M_0 + M_{\rm c} + \epsilon^\prime/c^2)
\end{eqnarray}
where $\epsilon^\prime$ is the dynamical efficiency measured in the
frame moving at $\beta_{\rm c}c$.  Solving for $\beta_{\rm c}$ and
$\epsilon^\prime$ gives
\begin{eqnarray}
\beta_{\rm c}\, &=&\, \beta_0 \, 
{ M_0 \Gamma_0 \over M_0 \Gamma_0 + M_{\rm c} }\, \equiv \,
 { \beta_0  \over 1+x}; \quad x\equiv {M_{\rm c} c^2 \over M_0\Gamma_0 c^2}
\label{betac}
\\
\epsilon^\prime \, &=&\, {E_0 \over \Gamma_{\rm c}} \, 
\left( 1+x - x\Gamma_{\rm c} - {\Gamma_{\rm c} \over \Gamma_0} \right) 
\\
\epsilon \, &\equiv &\Gamma_{\rm c}\epsilon^\prime. 
\end{eqnarray}
The Lorentz factor $\Gamma_{\rm c}$ is shown in Fig. \ref{tappo1} as a
function of $M_{\rm c}$ for four values of $E_0$ (top panel) and as a
function of $E_0$ for three values of $M_{\rm c}$ (bottom panel).
One can see that for the process to be interesting (i.e.
$\Gamma_{\rm c}$ significantly smaller than $\Gamma_0$)
and to avoid ``over--loading" of baryons (too small $\Gamma_{\rm c}$)
$M_{\rm c}$ is required to be in specific ranges, that depend
on $E_0$. These ranges, however, encompass almost two orders of magnitude.

\begin{figure}
\vskip -0.5 true cm
\centerline{\psfig{figure=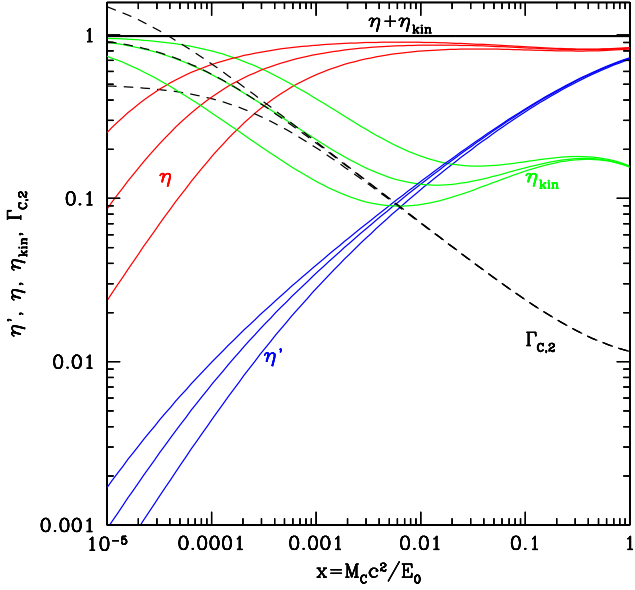,width=9.5cm,height=9.5cm}}
\vskip -0.7 true cm
\caption{The Lorentz factor $\Gamma_{\rm c,2}\equiv \Gamma_c /100$, 
the dynamical efficiency $\epsilon^\prime$ and $\epsilon$
(in the comoving and observer frame, respectively) as a function of $x$.  
Each quantity is calculated for $\Gamma_0=50$, 100 and 200 (from
top to bottom).}
\label{tappoeff}
\end{figure}

$\Gamma_{\rm c}$ as a function of $x$ is reported in
Fig. \ref{tappoeff}.  The power--law dependence $\Gamma_{\rm c}\propto
x^{-1/2}$ can be derived directly from Eq. \ref{betac}, since
$\Gamma_{\rm c}^{-2} = 1-\beta_{\rm c}^2 \propto 2x$ for $x \ll 1$ and
$\beta_0\to 1$. $\Gamma_{\rm c}$ is limited to $\sim 10$ even for
$x\sim 4\times 10^{-3}$, corresponding to $M_{\rm c} = 2.2\times
10^{-5} E_{0, 52} M_\odot$.

In Fig. \ref{tappoeff} we show $\eta^\prime\equiv \epsilon^\prime/E_0$ and
$\eta\equiv \epsilon/E_0$ as a function of $x$. For clarity also the
fraction of the initial kinetic energy preserved after the collision
\begin{equation}
\eta_{\rm kin} \, \equiv \, {(\Gamma_{\rm c}-1) 
(M_0+M_{\rm c} ) c^2 \over E_0} \, =\,
(\Gamma_{\rm c}-1) \left( x + {1\over \Gamma_0} \right)
\end{equation}
is plotted. The sum $\eta+\eta_{\rm kin}$ is unity by definition.

Note that since $x$ is a ratio, $M_{\rm c}$ and $E_0$ can be taken
both as ``isotropic" or ``real" (collimation corrected) values.

\subsection{Evolution of the fireball+IDM system}

After the collision/dissipation phase the fireball+IDM is expected to
be optically thick: the Thompson scattering optical depth of the IDM
material is
\begin{equation}
\tau_{\rm c} \, =\, \sigma_{\rm c} n \Delta R \, =\, {\sigma_{\rm T}
M_{\rm c} \over 4\pi R_{\rm c}^2 m_{\rm p}} \, =\, 3.2\times 10^3 \, {
M_{\rm c, -6} \over R_{\rm c, 11}^2},
\label{tauc}
\end{equation}
while the optical depth of the fireball just before the
collision is of order 
\begin{equation}
\tau_0 \, =\, 
{\sigma_{\rm T} E_0 \over 4\pi R^2 \Gamma_0 m_{\rm p} c^2} \, =\, 
1.8\times 10^5 \, { E_{\rm F, 52} \over R_{11}^2 },
\end{equation}
where $M_{\rm c}$ and $E_0$ are here isotropic quantities.  These
large optical depths imply that the radiation produced following the
collision is trapped inside the fireball+IDM system, which will expand
because of the internal pressure.  In the frame moving with
$\Gamma_{\rm c}$, the expansion is isotropic.  In this frame some
final $\Gamma^\prime$ will be reached.  As seen in the observer frame,
the expansion is highly asymmetric, and the geometry of the system
resembles a cone, with semi--aperture angle given by (Barbiellini,
Celotti \& Longo 2003)
\begin{equation}
\tan\theta_j \, =\, {\beta_\perp \over \beta_\parallel} \,=\,
{\beta^\prime \sin\theta^\prime \over 
\Gamma_{\rm c}(\beta^\prime \cos\theta^\prime +\beta_{\rm c}) } 
\,\, \to \,\,
\theta_j\, \sim \, {1\over \Gamma_{\rm c}},
\end{equation}
where the last equality assumed $\theta^\prime =90^\circ$
($\beta^\prime\sim 1$ and $\beta_{\rm c}\sim 1$).  Therefore the
aperture angle of the re-born fireball is related to $\Gamma_{\rm c}$,
independently of the initial $\Gamma$--factor of the fireball {\it
before} the collision or {\it after} the expansion.  Note that the
initial aperture angle of the fireball is irrelevant, as it is the
aperture angle of the funnel of the progenitor star.

\begin{figure}
\vskip -0.5 true cm
\centerline{\psfig{figure=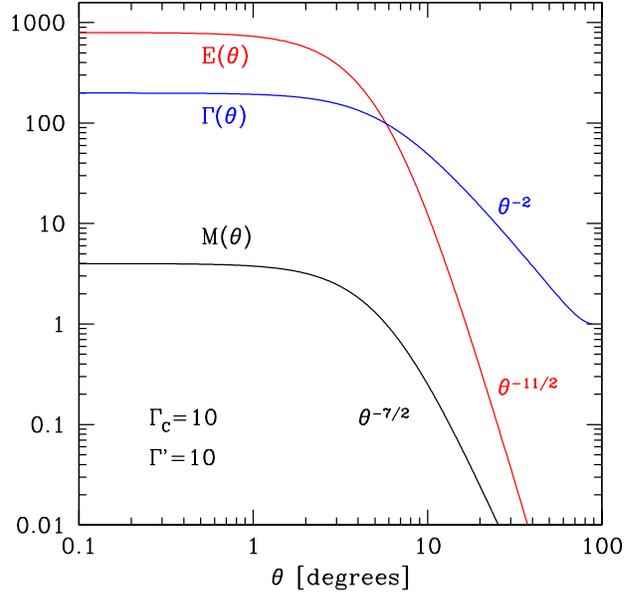,width=9.5cm,height=9.5cm}}
\vskip -0.7 true cm
\caption{The mass, energy and bulk Lorentz factor (observer frame) as
functions of the angle from the jet axis $\theta$, for $\Gamma_{\rm
c}=10$ and $\Gamma^\prime=10$.  All the three quantities are constant
up to $\theta \simeq 1/\Gamma_{\rm c}$ and then decrease
(approximately) as power--laws.}
\label{e}
\end{figure}

The fireball is not collimated in a perfect cone, and mass and energy
propagate also outside $\theta_j$.  Since in the frame moving with
$\Gamma_{\rm c}$ it expands isotropically, $M'(\Omega')=M'/(4\pi)$
is approximately constant.  Therefore in the observer frame
\begin{equation}
M(\theta) \, =\, {M'\over 4\pi} \, {d \cos\theta' \over d\cos\theta}.
\end{equation}
Also the resulting Lorentz factor $\Gamma$ is angle dependent: from
the relativistic composition of velocity (e.g. Rybicki \& Lightman
1979):
\begin{eqnarray}
\beta_\parallel \, &=&\, \beta\cos\theta \, = \, 
  { \beta^\prime \cos\theta^\prime +\beta_{\rm c} 
  \over (1+ \beta_{\rm c}\beta^\prime \cos \theta^\prime ) } \nonumber \\
\beta_\perp     \, &=&\, \beta\sin\theta \, = \, 
{\beta^\prime \sin\theta^\prime   
  \over  \Gamma_{\rm c} (1+ \beta_{\rm c}\beta^\prime \cos \theta^\prime ) } \nonumber \\
\beta(\theta)   \, &=& \, \left(\beta^2_\parallel + \beta^2_\perp \right)^{1/2}
  \nonumber \\
    \Gamma(\theta) \, &=& \, \left[ 1-\beta^2(\theta)\right]^{-1/2}\, =\,
\Gamma^\prime \Gamma_{\rm c} (1+\beta^\prime\beta_{\rm c}\cos\theta^\prime), 
\end{eqnarray}
The observed $\Gamma$--factor is constant up to angles
slightly smaller than $1/\Gamma_{\rm c}$, and decreases
as $\theta^{-2}$ above.

As a consequence of the angular dependence of mass and bulk Lorentz
factor, also the energy depends on $\theta$ as
\begin{equation}
E(\theta) \, =\, \Gamma(\theta) M(\theta) c^2.
\end{equation}
Such dependences of mass, $\Gamma$ and $E$ on polar angle are
illustrated in Fig. \ref{e}. 
The jet is structured and well
approximated by a top--hat jet: the energy profile is nearly constant
within an angle slightly smaller than $1/\Gamma_{\rm c}$, and at
larger angles decreases approximately as a steep power--law
$E(\theta)\propto \theta^{-11/2}$], since $M(\theta)\propto
\theta^{-7/2}$ and $\Gamma(\theta)\propto \theta^{-2}$ for $\theta\gg
1/\Gamma_{\rm c}$.  This particular behavior gives raise to an
afterglow light curve indistinguishable from a top--hat jet (see Rossi
et al. 2004).

\section{Discussion and conclusions}

The most appealing feature of the proposed model is the high
efficiency in re--converting the fireball kinetic energy into
internal pressure at a radius comparable to the radius of the
progenitor star, i.e. on a scale not far from the transparency one.
Also the observed energetics of the internal radiation will be
large, since the system becomes transparent during (or
slightly after) the expansion/acceleration phase,
similarly to the
standard fireball models of initially high entropy,
where the fireball becomes transparent before coasting.
Our model therefore increases the parameter space of 
high efficiency regimes.
Our model is also similar to
the model proposed by T06 and
T07 and to those in which the dissipation of an energetically
important magnetic field occurs at large radii (see e.g. Giannios \&
Spruit 2007). 
The efficiency of the energy re--conversion for the
fireball--IDM collision is of the order of 50--80 per cent for large
ranges in the mass of the IDM and energy of the fireball (see
Fig. \ref{tappoeff}).

With respect to the idea proposed by T07, summarised in \S 2, in our
scenario there is no requirement on any specific value for the
fireball bulk Lorentz factor $\Gamma_0$ prior to its collision with
the IDM.  Note also that in the T07 model, a ``standard" fireball
(i.e. not magnetic) moving with $\Gamma_0\sim 1/\theta_{\rm j}$ can
dissipate part of its kinetic energy, but it cannot reach final
$\Gamma$--factors larger than $\Gamma_0$, which is bound to be small 
for typical $\theta_{\rm j}$.  
In our case the final $\Gamma$ can instead be large, 
(even if always smaller than $\Gamma_0$).
The jet angle is $\sim 1/\Gamma_{\rm c}$ 
even for large values of the initial $\Gamma_0$ and final $\Gamma$.
This is a result of our model, and not an assumption.

The re--born fireball is structured and the $M(\theta)$ and
$\Gamma(\theta)$ behaviors imply that $E(\theta)$ depends on $\theta$
as steep power--laws.  Despite the angle dependence of the energy, the
jet should produce an afterglow indistinguishable from a top--hat jet.
Clearly our description of the fireball--IDM interaction is
extremely simplified, aimed at building a physical intuition based on
the analytical treatment.  More realistic situations should be studied
via numerical simulations, but we would like to comment on two
aspects. i) Even if the IDM is initially at rest, as soon as the
fireball starts depositing a fraction of energy and momentum, the IDM
would begin to move.  The whole process will take long enough that
towards the end it would be probably better described as the
interaction with a moving IDM, with a consequent loss of efficiency.
ii) In a `continuos' (non--intermittent) scenario, the IDM would
predominantly interact with the fireball edge, causing only a partial
dissipation of its energy and the formation of a ``fast spine--slow
layer" structure.  In this case the determination of the relevant jet
opening angle (i.e. within which most of the energy is concentrated)
requires a more accurate numerical treatment. 

Despite of these caveats, it is still interesting to consider whether
the model can account for the spectral--energy correlations, or at
least highlight the relations between them.  Consider the Ghirlanda
correlation in the wind case, $E_{\rm peak} \propto E_{\gamma}$, that
can be rewritten as $E_{\rm peak} \propto \theta_{\rm j}^2 E_{\rm
\gamma, iso}$ for small $\theta_{\rm j}$.  The requirement that also
$E_{\rm peak} \propto E^{1/2}_{\rm \gamma, iso}$ (Amati relation)
leads to:
\begin{equation}
E_{\gamma} \theta_{\rm j}^2 \, =\, {\rm constant}, 
\label{egtheta}
\end{equation}
i.e. more energetic bursts are more collimated, as
predicted in our model.
The above condition (Eq. \ref{egtheta}) can be quantitatively
satisfied if 
i) the mass against which the fireball collides is
similar in different GRBs, namely $\Gamma_{\rm c}\sim 1/ \theta_{\rm j} 
\sim E_0^{1/2}$ (see Fig. \ref{tappo1}) and 
ii) the prompt emission luminosity $E_\gamma$ is also a constant 
fraction of $E_0$ for different GRBs.
Eq. \ref{egtheta} does not explain the Amati or Ghirlanda relations, 
although it offers some physical meanings of required connection
between the two.
Note that, in the standard internal shock model model, one can
recover the Amati relation if $\Gamma\sim$ constant
(see Zhang \& Meszaros 2002).

In this Letter we did not discuss the characteristics of the spectrum
predicted in our scenario (Nava et al., in prep). In general terms,
the most effective radiation process would be ``dynamical Compton", or
Fermi ``acceleration" of photons, as discussed by Gruzinov \& Meszaros
(2000) in the context of internal shocks.  The large number of photons
per proton in the fireball (corresponding to the ``fossil" radiation
that has accelerated the fireball itself in the first place) implies
that a significant fraction of the internal energy following the
fireball--IDM collision {\it directly} energizes photons, i.e. photons
amplify their energy by interacting with leptons with bulk momentum
not yet randomized in the shock.  The process is analogous to particle
acceleration in shocks and gives raise to high energy photons,
conserving their number.

Since this occurs at $R\sim 10^{11}$ cm, i.e. slightly above the
progenitor star, the re--born fireball will become transparent during
or just after the acceleration phase (at a transparency radius
$R_\tau\sim 3\times 10^{12}$ cm, Eq. \ref{tauc}).  This ensures that
photons do not have time to loose energy via adiabatic expansion, once
again leading to large efficiencies.

The timescale for the refilling of the funnel during quiescent phases
of the central engine can be short enough that the process repeats
itself. The estimated refilling timescale is of the order of a second:
bursts with shorter or longer `quiescent' phases should then show
different properties.  If the fireball--IDM collision is the dominant
process for the dissipation (and emission), more energetic spikes are
expected to follow longer quiescent phases (see Ramirez--Ruiz \&
Merloni, 2001).

\section*{Acknowledgements}
A 2005 PRIN--INAF grant is acknowledged for partial funding.  We
thank the anonymous referee for useful criticism.

\end{document}